\documentclass[aps,twocolumn,showpacs,preprintnumbers,nofootinbib,
               amsmath,amssymb,superscriptaddress]{revtex4}
\usepackage{graphicx}

\newcommand{\feyn}[1]{
  \setbox0=\hbox{\ensuremath{#1}}
  \hbox to\wd0{\hbox to0pt{\hbox to\wd0{\hss/\hss}\hss}\box0}}
\newcommand{\MeV}{\,\text{MeV}}

\begin{document}

\title{Collective excitations in a superfluid of color-flavor locked
       quark matter}

\author{Kenji Fukushima}
\affiliation{Center for Theoretical Physics, Massachusetts Institute
 of Technology, Cambridge, Massachusetts 02139, USA}
\affiliation{Department of Physics, University of Tokyo,
 7-3-1 Hongo, Bunkyo-ku, Tokyo 113-0033, Japan}

\author{Kei Iida}
\affiliation{RIKEN BNL Research Center, Brookhaven National
 Laboratory, Upton, New York 11973-5000, USA}

\begin{abstract}
     We investigate collective excitations coupled with baryon density
in a system of massless three-flavor quarks in the collisionless
regime.  By using the Nambu--Jona-Lasinio (NJL) model in the
mean-field approximation, we field-theoretically derive the spectra
both for the normal and color-flavor locked (CFL) superfluid phases at
zero temperature.  In the normal phase, we obtain usual zero sound as
a low-lying collective mode in the particle-hole (vector) channel.  In
the CFL phase, the nature of collective excitations varies in a way
dependent on whether the excitation energy, $\omega$, is larger or
smaller than the threshold given by twice the pairing gap $\Delta$, at
which pair excitations with nonzero total momentum become allowed to
break up into two quasiparticles.  For $\omega\ll2\Delta$, a phonon
corresponding to fluctuations in the $U(1)$ phase of $\Delta$ appears
as a sharp peak in the particle-particle (``$H$'') channel.  We
reproduce the property known from low energy effective theories that
this mode propagates at a velocity of $v_H=1/\sqrt{3}$ in the low
momentum regime; the decay constant $f_H$ obtained in the NJL model is
identical with the QCD result obtained in the mean-field
approximation.  We also find that as the momentum of the phonon
increases, the excitation energy goes up and asymptotically approaches
$\omega=2\Delta$.  Above the threshold for pair excitations
($\omega>2\Delta$), zero sound manifests itself in the vector
channel.  By locating the zero sound pole of the vector propagator in
the complex energy plane we investigate the attenuation and energy
dispersion relation of zero sound.  In the long wavelength limit, the
phonon mode, the only low-lying excitation, has its spectral weight in
the $H$ channel alone, while the spectral function vanishes in the
vector channel.  This is due to nontrivial mixing between the $H$ and
vector channels in the superfluid medium.  We finally extend our study
to the case of nonzero temperature.  We demonstrate how Landau damping
smears the phonon peak in the finite temperature spectral function.
We find a pure imaginary pole of the $H$ propagator in the complex
energy plane, which can be identified as a diffusive mode responsible
for the Landau damping.  From the pole position we derive the thermal
diffusion constant.
\end{abstract}
\preprint{MIT-CTP-3576}
\maketitle

%%%%%%%%%%   Introduction   %%%%%%%%%%

\section{introduction}

     Quark matter at high baryon density is considered to have a rich
phase structure in a way dependent on the physical conditions for
color and flavor such as weak equilibrium and color neutrality.  For
the past few decades it has been predicted that cold quark matter is
color superconducting when the baryon density is sufficiently
high~\cite{reviews}.  Color superconductivity occurs in quark matter
in a similar manner to ordinary superconductivity in metals; any
attractive force between quarks can bring about an instability of the
Fermi surface against Cooper pairing, which subsequently leads to a
BCS superfluid state characterized by condensation of quark Cooper
pairs and to a color version of the Meissner effect through
nonvanishing color charge carried by the pairs.  In fact, the
quark-quark force induced by one-gluon exchange is attractive in the
color antisymmetric channel.  At asymptotically high densities where
the interactions are dominated by one-gluon exchange, this attractive
force should drive Cooper instability.  More recently, effective model
analyses, rather than rigorous QCD calculations, have predicted that
quark matter is a color superconductor even at several times normal
nuclear density and that the zero temperature pairing gap and thus the
critical temperature are of order 10--100$\MeV$.

     In possible physical realizations in heavy ion collisions and
neutron stars, color superconducting quark matter might leave
evidences for color superconductivity via precursory phenomena above
the critical temperature~\cite{Kitazawa:2001ft}, neutrino emission
from proto-neutron stars~\cite{Carter:2000xf}, and neutron star
cooling~\cite{Page:2000wt,Alford:2004zr}, etc.  To obtain such
possible evidences, it is indispensable to clarify the equilibrium
properties of color superconductors such as the equation of state and
elementary excitations.  This paper elaborates on collective
excitations in quark matter, which might help us detect the onset of
color superconductivity.

     Since quarks have more internal degrees of freedom than electrons
by two (color and flavor), various pairing patterns are possible among
quarks, even within a state of $S$-wave, spin-singlet pairing.  The
ratio between the strange quark mass and the quark chemical potential
is known to be a crucial parameter that determines the energetically
favorable pairing pattern.  In the case of three massless flavors the
color-flavor locked (CFL) phase~\cite{Alford:1998mk} is the ground
state~\cite{Schafer:1999fe}.  In this paper we shall focus only on the
massless limit of CFL quark matter in which the baryon number
conservation is spontaneously broken by the condensation of quark
pairs.  This can be mathematically formulated as global $U(1)_B$
symmetry breaking, as we briefly recapitulate in
Sec.~\ref{sec:cooper}.

     Superfluid $^3$He, a typical laboratory fermion system that
exhibits global $U(1)$ symmetry breaking~\cite{3He}, is similar to CFL
quark matter in many respects.  In fact, the ground state of liquid
$^3$He, i.e., the B phase, has a condensate {\em isotropic} in
spin-orbit space, which is similar to the CFL phase in which the
condensate is {\em isotropic} in color-flavor space.  In both systems,
responses to rotation are characterized by a lattice of
vortices~\cite{III}.

     In the B phase of superfluid $^3$He in which the pairing gap has
no nodes, thermal quasiparticle excitations are suppressed at
temperatures much lower than the gap.  In this case, the thermodynamic
properties such as the specific heat are presumably controlled by
phonon excitations, Nambu-Goldstone bosons corresponding to
fluctuations in the $U(1)$ phase of the pairing gap, which appear in
the low energy spectral function in the particle-particle channel.
Then,  the same kind of phonons in CFL matter (often denoted by $H$)
are expected to play a key role in the thermodynamic properties at low
temperatures~\cite{Reddy:2002xc,Kundu:2004mz}.  In this paper we
investigate the $H$ phonon spectrum in the energy range including the
scale beyond the pairing gap $\Delta$ in the random phase
approximation of a Nambu--Jona-Lasinio (NJL) model for CFL quark
matter.  This microscopic approach enables us to see that $H$, if
embedded in the quasiparticle continuum lying above $2\Delta$, would
decay into two quasiparticle quarks.

     We remark that an $H$ mode does not appear as a transverse mode,
a feature responsible for a nonzero value of the superfluid baryon
density at zero temperature~\cite{II}.  Transverse excitations will be
ignored in the present study, but play a role in breaking up phase
coherence in the superfluid state and hence are relevant for responses
to magnetic field and rotation~\cite{III,cflmag} and for reduction in
the transition temperature~\cite{Matsuura:2003md}.

     In normal quark matter we also describe zero sound as
fluctuations in the baryon density by analyzing the spectral function
in the particle-hole (vector) channel in the random phase
approximation in which collisions between quark quasiparticles are
ignored (see Ref.~\cite{Chin:1977iz} for general arguments of
relativistic zero sound).  We extend the description of zero sound to
the case of the CFL phase.  We find that in the vector channel, zero
sound manifests itself as an attenuated mode in the quasiparticle
continuum lying above twice the gap, while $H$ dominates the spectrum
below the continuum.  We remark that for general excitation momenta
the random phase approximation is strictly applicable to weakly
coupled systems.  In this respect, a recent observation of zero sound
in the lattice simulation based on a QCD-like
model~\cite{Hands:2003dh} is noteworthy.

     Note that in the collisionless regime the restoring force of the
collective modes is a self-consistent field formed by a number of
particles, which would be disrupted by interparticle collisions if
any.  Collective modes in the hydrodynamic regime such as first sound,
of which the restoring force is provided by the collisions, will be
addressed in our future study.  For the study of low-lying collective
modes in both regimes, the Landau theory of Fermi liquids is useful.
This is the case with normal liquid $^3$He~\cite{Landau:1957},
superfluid $^3$He~\cite{Wolfle}, and normal quark
matter~\cite{Baym:1975va}.

     In calculating the spectral functions in the CFL phase, it is
important to take into account the mixing between the
particle-particle and particle-hole channels.  This is because such
mixing plays a role in determining the spectral weight of $H$ and zero
sound modes.  We find that in the long wavelength limit the spectral
function in the particle-hole channel vanishes due to the mixing
effect.

     We summarize the main conclusions of this paper as follows:

\begin{itemize}

\item   The velocity of the $H$ mode in the small momentum regime is
$1/\sqrt{3}$ at zero temperature.  As the momentum goes up, the mode
energy monotonically approaches $2\Delta$, above which individual pair
excitations are kinematically allowed.  The velocity $1/\sqrt{3}$
corresponds to the ratio of the spatial and temporal components of the
decay constant $f_H$.  We find that the expression for $f_H$ in the
NJL model adopted here is the same as the QCD result in the mean-field
approximation.

\item   In the CFL phase zero sound appears as an attenuated mode in
the quasiparticle continuum extending above $2\Delta$.  The
attenuation width decreases with increasing momentum.

\item   In the CFL phase, the spectral function in the vector channel
is strongly suppressed at small momenta by the mixing with the
particle-particle channel.

\item   At finite temperature at which thermally excited
quasiparticles are present, the $H$ mode has a finite decay width due
to Landau damping.  Correspondingly, a pure imaginary pole appears in
the $H$ propagator.  We estimate the thermal diffusion constant from
the pole behavior in the complex energy plane.

\end{itemize}

     We note that the present work, which focuses on collective modes
coupled with baryon density fluctuations in the CFL phase, is a
complement to the work by Gusynin and Shovkovy~\cite{Gusynin:2001tt}
that gave a detailed account of collective modes coupled with color
current fluctuations in the CFL phase.

     This paper is organized as follows:  We briefly review the CFL
Cooper pairing and the symmetry breaking pattern in
Sec.~\ref{sec:cooper}.  We describe a model for quark matter, the
self-consistency equations, and the spectral functions in
Sec.~\ref{sec:model}.  Full expressions for the spectral functions are
listed in Appendix~\ref{app:explicit}.  Section~\ref{sec:zero} is
devoted to showing the results for $H$ and zero sound modes at zero
temperature.  In Sec.~\ref{sec:finite}, we present the results at
finite temperature and discuss Landau damping of the $H$ mode.
Concluding remarks are given in Sec.~\ref{sec:conclusion}.
Appendix~\ref{app:analytic} explains how we perform the analytic
continuation of the propagators of collective excitations to search
for the poles in the complex energy plane.  We work in units
$\hbar=c=k_B=1$.

%%%%%%%%%%   Cooper Pair, Symmetry Breaking, and Nambu-Goldstone Bosons   %%%%%%%%%%

\section{Cooper pairing, symmetry breaking, and Nambu-Goldstone bosons}
\label{sec:cooper}

     In this section, we summarize the fundamental properties of the
CFL phase.  In particular, the symmetry breaking pattern and the
associated Nambu-Goldstone bosons are mentioned.

     In the CFL phase with $uds$-flavor and $RGB$-color massless
quarks, the energetically favored Cooper pairing of quarks is
antisymmetric in color space and has even parity and zero total
angular momentum~\cite{Alford:1998mk,Schafer:1999fe}.  Then, the
antisymmetry of the pairing in flavor space follows from the Pauli
principle.  [Note that in the absence of quark masses, only quarks of
the same chirality can be paired~\cite{Pisarski:1999av}.]  The
corresponding diquark condensate reads
\begin{equation}
 \langle\bar{\psi}^C_{\alpha i}\gamma_5\psi_{\beta j}\rangle
  \sim \Delta\;(\delta_{\alpha i}\delta_{\beta j}
  -\delta_{\alpha j}\delta_{\beta i}),
\label{eq:condensate}
\end{equation}
where the Roman and Greek subscripts stand for the indices in flavor
and color space, respectively, $\psi$ is the quark spinor, and
$\psi^C=C\bar{\psi}^T$ is the charge conjugate spinor with the charge
conjugation matrix $C=i\gamma^2\gamma^0$ in the Pauli-Dirac
representation.  A precise relation between the condensate and the
pairing gap will be given in Eq.~(\ref{eq:delta}) after introducing a
model for quark matter.  The coexisting condensate in the CFL phase,
which arises in a color and flavor symmetric state, will be ignored
since it makes negligible difference in the present analysis that
focuses on the superfluid properties.

     The condensate~(\ref{eq:condensate}) breaks both $SU(3)_L$ and
$SU(3)_R$ of chiral symmetry, whereas it leaves unbroken a symmetry
under  a vector rotation in flavor space and a simultaneous color
rotation in the opposite orientation.  The symmetry breaking pattern
in the CFL phase can thus be written as~\cite{reviews,Alford:1998mk}
\begin{equation*}
 \begin{split}
 & [SU(3)_{\text{color}}] \times
  SU(3)_L \times SU(3)_R \times U(1)_B \\
 & \to SU(3)_{C+L+R} \times \mathbb{Z}_2 \,.
 \end{split}
\end{equation*}
Here, $\mathbb{Z}_2$ is a symmetry under $\psi\to-\psi$.  Nine
Nambu-Goldstone bosons result from the spontaneous symmetry breaking
described above.  Eight pion-like bosons associated with the breakdown
of the axial part of chiral symmetry form an octet in $SU(3)_{C+L+R}$
just like $(\pi,K,\eta)$ in hadronic matter.  One more massless boson
arises as a color-flavor singlet from the spontaneous breaking of
baryon number conservation in a manner that preserves local electric
and color charge neutrality.  This singlet boson is a phonon (often
referred to as $H$) corresponding to $U(1)_B$ phase fluctuations of
the condensate~(\ref{eq:condensate}).

     For completeness, let us consider the other phases which are
predicted to appear in the phase diagram for neutral quark matter with
nonzero strange quark
mass~\cite{Iida:2003cc,Ruster:2004eg,Fukushima:2004zq} and confirm
that all these phases do not have an $H$ mode.  It is well known that
in the 2SC phase in which only two colors and two flavors participate
in pairing, there remains an unbroken baryon number symmetry
associated with the unpaired color and flavor.  In the same way, the
uSC and dSC phases in which only $ds$ and $us$ pairing is breached as
compared with the pairings in the CFL phase, respectively, preserve a
modified $U(1)_B$ symmetry.  The uSC phase, for instance, has finite
condensates $\langle ud\rangle$ and $\langle su\rangle$.  These two
condensates are invariant simultaneously under rotations generated by
an appropriate linear combination of the $U(1)_B$ generator and two
$U(1)$ generators corresponding to flavor number conservation.
Therefore, there is no $H$ in the 2SC, uSC, and dSC phases.  Put
another way, in each of these phases, $U(1)_B$ phase fluctuations in
the order parameter are inevitably coupled with electric and color
charge density since the condensates, as a whole, are not charge
neutral in electricity and color.  We note that in the CFL phase
modified by quark masses different among flavors, $H$ modes could be
affected by coupling with electric and color charge density.  In this
case, the quark system, if being color neutral and being neutralized
and $\beta$-equilibrated by an electron gas, would generally have a
superfluid part that is no longer charge neutral in electricity and/or
color.  This nonneutral superfluid part would occur not only at
nonzero temperature, but also in the presence of gapless quark modes
even at zero temperature~\cite{Alford:2003fq}.

\begin{figure}
 \includegraphics[width=4cm]{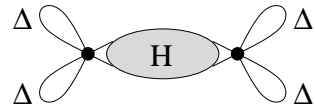}
 \caption{Schematic picture of the $H$ propagation.  Colored diquark
 condensates supply color charge in such a way as to make $H$ a color
 singlet at the edges.}
 \label{fig:h_prop}
\end{figure}

     All the Nambu-Goldstone bosons in the CFL
condensate~(\ref{eq:condensate}) are color singlets because the
corresponding fluctuations in the order parameter are not coupled with
color charge density and thus are not eaten by the longitudinal
component of color gauge fields through the Anderson-Higgs mechanism.
It follows that the CFL pions and $H$ consist of at least four and six
quarks, respectively~\cite{Fukushima:2004bj}, i.e., they are composed
of such combinations of particle-particle and hole-hole pairs as
$(RG)(\bar{R}\bar{G})$ and $(RG)(GB)(BR)$.  In this paper we assume
that colored diquark condensates compensate color charge at the edges
of the $H$ propagator as shown in Fig.~\ref{fig:h_prop}.  We can then
describe the propagating part of $H$ by a state having an excited
quark Cooper pair in Fock space.  This picture is based on a
mean-field approximation, which is valid as long as the size of the
mean-field $\Delta$ is larger than the size of quantum and thermal
fluctuations around $\Delta$.

%%%%%%%%%%   Model Calculations   %%%%%%%%%%

\section{model calculations}
\label{sec:model}

     In this section we first describe the equilibrium properties of a
normal fluid and a superfluid of quark matter with three massless
flavors by utilizing an effective model with local four-quark
interactions.  We then examine excitation spectra in both phases in
the random phase approximation.

%---   Self-consistency equations   ---%

\subsection{Self-consistency equations}

     We begin with a model for massless three-flavor quark matter.
For the later purpose of describing collective excitations in the
particle-particle ($H$) and particle-hole (vector) channels, it is
convenient to deal with the Lagrangian density having the diquark and
vector interactions.  One can generally obtain such interaction terms
from any local four-quark interactions after appropriate Fierz
transformation.  Then the Lagrangian density can be written as
\begin{equation}
 \mathcal{L} = \bar{\psi}(i\feyn{\partial}+\mu_q \gamma^0)\psi
  +\mathcal{L}_D + \mathcal{L}_V,
\label{eq:interaction}
\end{equation}
where $\mu_q$ is the quark chemical potential, $\mathcal{L}_D$ and
$\mathcal{L}_V$ are the diquark and vector interaction terms as will
be specified below.  As we will see later, $\mu_q$ receives a finite
correction from the vector mean-field.  Hereafter we will take
$\mu_q=500\MeV$.  The corresponding baryon density is, as we will see
later, approximately $10\rho_0$ in our calculation where
$\rho_0\simeq0.16\,\text{fm}^{-3}$ is the normal nuclear density, and
may be relevant in the cores of compact stellar objects, although
ideally symmetric CFL quark matter as considered here is only relevant
at much higher densities where the strange quark mass is negligibly
small compared with the quark chemical potential.  Note, however, that
we have technical difficulty in taking an extremely large value of
$\mu_q$, because the effective model with the interaction terms that
will be shown below is non-renormalizable and needs a finite cut-off
parameter, $\Lambda$.  We shall choose $\Lambda=1\,\text{GeV}$ in this
paper.  Then, our choice, $\mu_q=500\MeV$, can be considered
effectively as the upper limit at which quark excitations around the
Fermi surface would hardly suffer from any cut-off artifact.  We thus
expect that our analysis of collective excitations with momenta up to
$\sim100\MeV$ is free from such artifact.

     In this paper, for simplicity, we choose $\mathcal{L}_D$ to be
the interaction that occurs only in the color-flavor singlet diquark
channel.  This ansatz for $\mathcal{L}_D$ makes no difference in the
resulting gap equations as far as ideally symmetric CFL quark matter
is concerned.  The four-quark interactions thus take the following
forms:
\begin{align}
 \mathcal{L}_D &= \frac{G}{12}(\bar{\psi}_{\alpha i}
   P_{\alpha\beta}^{ij}\psi^C_{\beta j}) (\bar{\psi}^C_{\alpha'i'}
   P_{\alpha'\beta'}^{i'j'}\psi_{\beta'j'}) \,,\\
 \mathcal{L}_V &= -\frac{G_V}{2}(\bar{\psi}\gamma_\mu\psi)
   (\bar{\psi}\gamma^\mu\psi) \,,
\end{align}
with $P_{\alpha\beta}^{ij}=i\gamma_5\epsilon^{ijk}
\epsilon_{\alpha\beta k}$ in which the sum is taken over $k$.

     The value of $G_V$ is of the order of $G$, but our qualitative
results for elementary excitations are independent of a particular
choice of $G_V$.  Hereafter we shall take $G_V=G/2$ for definiteness.
A different $G_V$ provides a different (dimensionless) spectral weight
in the vector channel, which is approximately scaled by $G_V$ unless
$G_V$ is anomalously large.

     It should be emphasized that the vector interaction generally
arises since chiral symmetry is intact; $\mathcal{L}_V$ is chirally
symmetric itself.  For specific values of $G_V$ the phase diagram on
the temperature versus chemical potential plane can have such
intricate structures as to contain multiple critical end-points
associated with diquark and chiral
condensates~\cite{Kitazawa:2002bc}.  Since the density region of
interest here is presumably beyond the chiral transition density, we
simply ignore effects of chiral condensation.

     As to the diquark interaction, we shall fix $G$ in such a way
that the solution to the self-consistency equations with $G_V=G/2$
yields $\Delta=25\MeV$ with $\mu_q=500\MeV$ at zero temperature.  As
we will see in Sec.~\ref{sec:zero}, the properties of collective
excitations in the CFL state change drastically according to whether
they reside above or below $2\Delta$, which corresponds to a threshold
above which pair excitations are allowed to occur.  As long as
$\Delta$ is much smaller than the quark chemical potential and much
larger than the temperature, the energy and momentum of collective
excitations are essentially scaled by $\Delta$.  In this case, a
single example presented here with the choice $\Delta=25\MeV$ suffices
for us to deduce what takes place for general values of $\Delta$.

     Let us proceed to adopt the mean-field approximation by
introducing variational variables, i.e., the gap parameter, $\Delta$,
and the quark number density, $n_q$, as
\begin{align}
 \Delta &= \frac{G}{6}\langle\bar{\psi}_{i\alpha}P_{\alpha\beta}^{ij}
  \psi^C_{j\beta}\rangle ,
\label{eq:delta}\\
 n_q &= \langle\bar{\psi}\gamma^0\psi\rangle .
\label{eq:rho}
\end{align}
As we shall see, we obtain the self-consistency equations that
determine these variables by substituting the mean-field quark
propagator with $\Delta$ and $n_q$ into the right side of
Eqs.~(\ref{eq:delta}) and (\ref{eq:rho}).  Equivalently, these
self-consistency equations can be obtained from the  stationary
conditions on the mean-field thermodynamic potential with respect to
$\Delta$ and $n_q$.  The corresponding equations are the gap equation
and the relation between the quark density and chemical potential.

     It is well known that the mean-field quark propagator in the CFL
phase takes the simplest form in the CFL basis,
$\psi_{\alpha i}=(\lambda^A/\sqrt{2})_{\alpha i}\psi^A$ and
$\bar{\psi}_{\alpha i}=\bar{\psi}^A(\lambda^A/\sqrt{2})_{i\alpha}$.
Here, the matrices, $\lambda^A$, in color-flavor space are the
Gell-Mann matrices for $A=1,\dots,8$ and
$(\lambda^0)_{i\alpha}=\sqrt{2/3}\delta_{i\alpha}$ for $A=0$, which
are normalized as
$\frac{1}{2}\text{tr}\lambda^A\lambda^B=\delta^{AB}$.  In the CFL
basis, the gap matrix in the quark propagator,
$\Delta P_{\alpha\beta}^{ij}$, has only diagonal components, i.e.,
$i\gamma_5\text{diag}\{\Delta_A\}$, where $\Delta_A=-\Delta$ for an
octet in color-flavor space ($A=1,\dots,8$) and $\Delta_A=2\Delta$ for
a singlet ($A=0$).

     In the present analysis of the self-consistency equations, one is
not allowed to simplify calculations by dropping the antiparticle part
in the quark propagator.  This is because the cancellation between the
particle and antiparticle contributions for momenta above the Fermi
surface is crucial for the evaluation of $n_q$ in the NJL model.  At
zero temperature, as a first approximation, one can estimate $n_q$
from the number density in a free quark gas as
$3\mu_q^3/\pi^2=4.9\,\text{fm}^{-3}$ for $\mu_q=500\MeV$, which
corresponds to a baryon density of roughly $10\rho_0$ as we have
mentioned before.  The NJL model calculation including both the
particle and antiparticle contributions results in
$n_q=5.1\,\text{fm}^{-3}$ for $\mu_q=500\MeV$ with $\Delta=25\MeV$.
This is consistent with the first estimate, while we obtain
$n_q=-14.7\,\text{fm}^{-3}$ in the absence of the antiparticle
contribution; even the sign is inconsistent.

     It is a usual technique to decompose the quark propagator into
the particle and antiparticle parts by using the energy projection
operators of noninteracting massless quarks,
$\Lambda^\pm_p=\frac{1}{2}(1\pm\gamma^0\vec{\gamma}\cdot\widehat{p})$,
where $\widehat{p}=\vec{p}/p$.  Using these operators, we can
explicitly write down the mean-field quark propagator in $2\times2$
Nambu-Gor'kov space as
\begin{equation}
 S_{AB}(p^\mu) =\delta_{AB}\gamma^0 \Biggl\{
  \left[\begin{array}{cc}
   \Lambda^-_p & 0 \\ 0 & \Lambda^+_p
  \end{array}\right] S^{\text{p}}_A
 \!+\! \left[\begin{array}{cc}
   \Lambda^+_p & 0 \\ 0 & \Lambda^-_p
  \end{array}\right] S^{\text{a}}_A \Biggr\} ,
\label{eq:propagator}
\end{equation}
where the particle part is
\begin{equation}
 S^{\text{p}}_A (p^\mu) = \frac{i}{p_0^2\!-\!(\epsilon^-_{\Delta_A})^2}
  \left[\begin{array}{cc}
    p_0+(p\!-\!\mu_r) & -i\Delta_A \gamma_5\gamma^0 \\
    -i\Delta_A \gamma_5\gamma^0 & p_0-(p\!-\!\mu_r)
   \end{array}\right] ,
\label{eq:propagator_p}
\end{equation}
the antiparticle part is
\begin{equation}
 S^{\text{a}}_A (p^\mu) = \frac{i}{p_0^2\!-\!(\epsilon^+_{\Delta_A})^2}
  \left[\begin{array}{cc}
    p_0-(p\!+\!\mu_r) & -i\Delta_A \gamma_5\gamma^0 \\
    -i\Delta_A \gamma_5\gamma^0 & p_0+(p\!+\!\mu_r)
   \end{array}\right] ,
\label{eq:propagator_a}
\end{equation}
and we have chosen the Nambu-Gor'kov basis as
\begin{equation}
 \bar{\Psi}(p^\mu)
  = \bigl[\bar{\psi}(p^\mu), \bar{\psi}^C (-p^\mu)\bigr],\quad
 \Psi(p^\mu)= \biggl[
  \begin{array}{c}\psi(p^\mu) \\ \psi^C (-p^\mu) \end{array}
  \biggr] \,.
\end{equation}
Here we have defined the energy of quark and antiquark quasiparticles
as $\epsilon^\pm_{\Delta_A}=\sqrt{(p\pm\mu_r)^2+\Delta_A^2}$ and the
quark chemical potential renormalized by the vector interaction term
as
\begin{equation}
 \mu_r=\mu_q-G_V \cdot n_q \,,
\label{eq:renom_chem}
\end{equation}
which amounts to $\mu_r=434.7\MeV$ for our parameter choice at zero
temperature.

     We can finally obtain the self-consistency equations at given
temperature $T$ by substituting the off-diagonal and diagonal
components of $S_{AB}$ given by Eq.~(\ref{eq:propagator}) into the
right side of Eqs.~(\ref{eq:delta}) and (\ref{eq:rho}), respectively.
Here we set $p_0=in\pi T$ and take the Matsubara frequency sum over
odd $n$.  The results read
\begin{align}
 1-\frac{G}{3\pi^2}\int^\Lambda \!\!dp\,p^2\Biggl[ &\frac{2}
  {\epsilon^-_\Delta}\tanh\biggl(\frac{\epsilon^-_\Delta}{2T}\biggr)
  +\frac{1}{\epsilon^-_{2\Delta}}\tanh\biggl(\frac{
  \epsilon^-_{2\Delta}}{2T}\biggr) \notag\\
 & +\text{(same with $\epsilon^- \to \epsilon^+$)}\Biggr]=0 \,,
\label{eq:gap_eq} \\
 n_q\!-\!\frac{1}{2\pi^2}\int^\Lambda \!\!\!dp\,p^2\Biggl[ &
  \frac{8\partial\epsilon^-_\Delta}{\partial\mu_r}\tanh\biggl(\!
  \frac{\epsilon^-_\Delta}{2T}\!\biggr)
  \!+\!\frac{\partial\epsilon^-_{2\Delta}}{\partial\mu_r}\tanh
  \biggl(\! \frac{\epsilon^-_{2\Delta}}{2T}\!\biggr) \notag\\
 & +\text{(same with $\epsilon^- \to \epsilon^+$)}\Biggr]=0 \,.
\label{eq:chem_eq}
\end{align}
The first term in the square brackets is the particle contribution
from octet quarks; the second from singlet quarks.  The antiparticle
contributions lead to the same expressions as the particle
contributions with $\epsilon^-$ replaced by $\epsilon^+$.

    We determine $G$ and $\mu_r$ from Eqs.~(\ref{eq:renom_chem}),
(\ref{eq:gap_eq}), and (\ref{eq:chem_eq}) in such a way that the
solutions to these equations yield $\Delta=25\MeV$ with
$\mu_q=500\MeV$ at zero temperature, as we have explained above.  Once
we fix the model parameters at zero temperature, we can derive the
temperature dependence of $\Delta$ and $\mu_r$ from the same
equations.  Such self-consistency equations predict a second-order
phase transition from the CFL phase to unpaired quark matter at
$T_{\rm c}=18.15\MeV$.  This value of $T_{\rm c}$ is larger than the
BCS value of $\simeq0.57\Delta$ due to the two-gap structure of the
quark excitations ($-\Delta$ and $2\Delta$ for octet and singlet
quarks, respectively)~\cite{Schmitt:2002sc}.  We note, however, that
in a situation in which CFL quark matter behaves as a type-I color
superconductor, thermal fluctuations in the gauge fields could  play a
role in changing the phase transition from second to first order and
in lowering the transition temperature~\cite{Matsuura:2003md}.  The
analysis of this situation is beyond the scope of the present paper.
In Sec.~\ref{sec:finite}, we will present the finite temperature
results showing an appreciable Landau damping effect on $H$ modes at
$T=0.8T_{\rm c}$, under the expectation that CFL quark matter at
$T=0.8T_{\rm c}$ is not affected severely by thermally fluctuating
gauge fields.

%---   Collective excitations   ---%

\subsection{Collective excitations}

     We proceed to describe longitudinal collective excitations in a
superfluid as well as in a normal fluid of quark matter in the random
phase approximation.  The collective modes that we can describe in
this approximation are an $H$ phonon in the particle-particle channel
and zero sound in the particle-hole channel, and they lie in the
collisionless regime.

\begin{figure}
 \includegraphics[width=8cm]{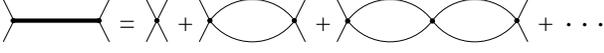}
 \caption{Schematic diagrams representing the propagation of
 collective excitations.}
 \label{fig:diagram}
\end{figure}

     In the random phase approximation, as we have briefly noted in
Sec.~\ref{sec:cooper}, we can assume that for $H$ modes, color
singletness is achieved at the edges of the propagator (see
Fig.~\ref{fig:h_prop}).  We can thus regard such collective
excitations as an excited Cooper pair with nonzero total momentum that
multiply scatter with each other in a superfluid medium, instead of
considering the propagation of three excited Cooper pairs.  We can
likewise consider zero sound in terms of the propagation of a quark
particle-hole pair both in a normal and a superfluid medium.
Consequently, the propagator of collective excitations can be
expressed as a sum over the bubble diagrams in the corresponding
channel as exhibited in Fig.~\ref{fig:diagram}.  In this subsection we
construct the vertices from the interaction terms in the $H$ and
vector channels and compute the bubble diagrams.

     Let us now construct matrices, $\Gamma$, at the vertices from the
interaction terms.  In doing so, we first rewrite the four-quark
interactions in terms of the Nambu-Gor'kov basis.  This is
straightforward although we should pay attention to how to count the
Nambu-Gor'kov replicas.  Since collective excitations couple not only
with the original fields but also with the replica fields, the
four-quark interactions in terms of the Nambu-Gor'kov basis involve
the original ones, the replica copies, and the original-replica cross
terms.  The resultant diquark interaction is
\begin{align}
 \mathcal{L}_D &\to 4\times\mathcal{L}_D \notag\\
 &= \frac{G}{12}\Bigl[\bigl(\bar{\psi}_{\alpha i}
  P_{\alpha\beta}^{ij}\psi^C_{\beta j}+\bar{\psi}^C_{\alpha i}
  P_{\alpha\beta}^{ij}\psi_{\beta j}\bigr)^2 \notag\\
 &\qquad\qquad
  -\bigl(\bar{\psi}_{\alpha i}P_{\alpha\beta}^{ij}\psi^C_{\beta j}
  -\bar{\psi}^C_{\alpha i}P_{\alpha\beta}^{ij}\psi_{\beta j}\bigr)^2
  \Bigr] \notag\\
 &= \frac{G}{12}\bigl[\bigl(\bar{\Psi}_{\alpha i}
  \Gamma_{D\alpha\beta}^{+ij}\Psi_{\beta j}\bigr)^2
  +\bigl(\bar{\Psi}_{\alpha i}\Gamma_{D\alpha\beta}^{-ij}
  \Psi_{\beta j}\bigr)^2\bigr],
\end{align}
where 
\begin{equation}
 \Gamma_{D\alpha\beta}^{+ij} \!=\!\biggl[ \begin{array}{cc}
  0 & P_{\alpha\beta}^{ij} \\ P_{\alpha\beta}^{ij} & 0
  \end{array} \biggr], \quad
 \Gamma_{D\alpha\beta}^{-ij} \!=\!\biggl[ \begin{array}{cc}
  0 & iP_{\alpha\beta}^{ij} \\ -iP_{\alpha\beta}^{ij} & 0
  \end{array} \biggr]
\label{eq:vertex_h}
\end{equation}
are the matrices associated with fluctuations in the amplitude and the
$U(1)$ phase of the diquark condensate, respectively.  We can likewise
express the vector interaction as
\begin{equation}
 \mathcal{L}_V \to 4\times\mathcal{L}_V
  = -\frac{G_V}{2}\bigl(\bar{\Psi}\Gamma_{V\mu} \Psi\bigr)
   \bigl(\bar{\Psi}\Gamma_V^\mu \Psi\bigr)
\end{equation}
with the matrix defined by
\begin{equation}
 \Gamma^\mu_V = \biggl[ \begin{array}{cc}
  \gamma^\mu & 0 \\ 0 & -\gamma^\mu
  \end{array} \biggr] \,.
\label{eq:vertex_v}
\end{equation}

     By using the quark propagator~(\ref{eq:propagator}) with the
definitions~(\ref{eq:propagator_p}) and (\ref{eq:propagator_a}) and
the matrices (\ref{eq:vertex_h}) and (\ref{eq:vertex_v}), we can
express the bubble diagrams as
\begin{align}
 &\Pi_H(q_0,\vec{q}) \notag \\
 &=-\frac{i}{2}\int^T \!\!\frac{d^4 p}{(2\pi)^4}\;
  \text{tr} \Gamma_D^- S(p_0\!+\!q_0,\vec{p}\!+\!\vec{q})\Gamma_D^-
  S(p_0,\vec{p})\,, \\
 &\Pi_V^{\mu\nu}(q_0,\vec{q})  \notag \\
 &=-\frac{i}{2}\int^T \!\!\frac{d^4 p}{(2\pi)^4}\;
  \text{tr} \Gamma^\mu_V S(p_0\!+\!q_0,\vec{p}\!+\!\vec{q})
  \Gamma^\nu_V S(p_0,\vec{p})\,, \\
 &\Pi^\mu_M(q_0,\vec{q}) \notag \\
 &=-\frac{i}{2}\int^T \!\!\frac{d^4 p}{(2\pi)^4}\;
  \text{tr} \Gamma_D^- S(p_0\!+\!q_0,\vec{p}\!+\!\vec{q})\Gamma^\mu_V
  S(p_0,\vec{p})
\end{align}
for the $H$, vector, and $H$-vector mixed channels, respectively.
Here, $\int^T d^4p/(2\pi)^4$ denotes the integration over $\vec{p}$
with the cut-off $\Lambda$ and the Matsubara summation with respect to
$p_0$, and the trace is taken over color, flavor, spinor, and
Nambu-Gor'kov indices.  The factor $1/2$ is necessary for adjusting
the duplicate loop counting in the Nambu-Gor'kov basis.

     In the present study we do not consider the bubble diagrams
including the vertex $\Gamma_D^+$ associated with fluctuations in the
amplitude of the order parameter.  This is partly because collective
amplitude modes are embedded in the quasiparticle continuum lying
above $2\Delta$ and heavily damped and partly because the important
decay process of an amplitude mode into two $H$ modes is beyond the
reach of the random phase approximation.

     We shall pick up only the $\mu=\nu=0$ component of the bubble
diagrams involved in the vector interaction, i.e., $\Pi_V^{00}$ and
$\Pi_M^0$, while discarding the longitudinal spatial components of the
bubble diagrams whose cut-off dependence is not well controllable in
the NJL model [see discussions around Eq.~(\ref{eq:exp_v}) for
details.]  These spatial components, even if taken into account, would
only change zero sound from superluminal to subluminal and would not
essentially change the $H$-vector mixing and the behavior of
attenuated zero sound in the CFL phase, as we shall see in the next
section.

     We list the explicit expressions for $\Pi_H$, $\Pi_V^{00}$, and
$\Pi_M^0$ in Appendix~\ref{app:explicit}.  We shall omit the
superscript ``0'' hereafter.

     It is convenient to arrange the polarizations and the coupling
constants in a matrix form
\begin{equation}
 \boldsymbol{\Pi}=\biggl[ \begin{array}{cc}
  \Pi_H & \Pi_M \\ -\Pi_M & \Pi_V
  \end{array} \biggr], \quad
 \boldsymbol{G}=\biggl[ \begin{array}{cc}
  \frac{G}{12} & 0 \\ 0 & -\frac{G_V}{2}
  \end{array} \biggr].
\end{equation}
It is clear from this form that we can analyze the $H$-vector mixing
effect by picking up eigenmodes in the corresponding channel.  It
should be noted that we always mean by the $H$ channel the channel
corresponding to phase fluctuations of the Cooper pair and  that $H$
as a physical excitation appears in an eigen-channel given by a
mixture of the $H$ channel and the vector channel.

     The retarded propagator as sketched in Fig.~\ref{fig:diagram} is
expressed in terms of $\boldsymbol{\Pi}$ and $\boldsymbol{G}$ as
\begin{equation}
 \boldsymbol{D}^{\text{R}}(\omega,\vec{q}) = \frac{1}
  {(2\boldsymbol{G})^{-1} - \boldsymbol{\Pi}(\omega+0^+,\vec{q})} .
\label{eq:ret_prop}
\end{equation}
This propagator contains all information on the dynamics of collective
excitations in the collisionless regime such as the mass, the
attenuation width, and the energy dispersion relation.  As a simple
exemplification, let us briefly make sure how the
propagator~(\ref{eq:ret_prop}) is consistent with the presence of a
massless $H$ mode.  In the limit of $\vec{q}=0$ and $\omega\to0$, the
mixed diagram $\Pi_M$ vanishes and the denominator of the $H$
propagator reduces to $6/G-\Pi_H(0,0)$.  Note that this denominator is
proportional to the left side of the gap equation~(\ref{eq:gap_eq}).
We thus find that it is zero given the explicit
expression~(\ref{eq:exp_h}).  This means that the
propagator~(\ref{eq:ret_prop}) has a massless pole in the $H$ channel
corresponding to the Nambu-Goldstone boson.

     For the purpose of clarifying the overall properties of
collective excitations, it is instructive to investigate the spectral
function, which is defined as
\begin{equation}
 \boldsymbol{\rho}(\omega,\vec{q})= -\frac{1}{\pi}\text{Im}
  \boldsymbol{D}^{\text{R}}(\omega,\vec{q})\,.
\label{eq:spectral}
\end{equation}
In the subsequent sections we will calculate the dimensionless
spectral functions in the $H$ and vector channels:
\begin{equation}
  \rho_H = \frac{6}{G}\bigl[\boldsymbol{\rho}
   \bigr]_{HH},\quad
  \rho_V = \frac{1}{G_V}\bigl[\boldsymbol{\rho}
   \bigr]_{VV},
\end{equation}
where $[\boldsymbol{\rho}]_{HH}$ is the $H$-$H$ component of the
$2\times2$ spectral matrix $\boldsymbol{\rho}$ and
$[\boldsymbol{\rho}]_{VV}$ is the $V$-$V$ component.  We will then
identify the poles of the propagator in the complex energy plane that
characterize the spectral shape.

%%%%%%%%%%   Zero Temperature Results   %%%%%%%%%%

\section{zero temperature results}
\label{sec:zero}

     In this section, we analyze the spectral functions at zero
temperature.  We clarify the properties of $H$ phonons in the CFL
phase and of zero sound in the normal and CFL phases.

%---   Phonon in a superfluid   ---%

\subsection{Phonon in a superfluid}

     We start with an $H$ phonon in the CFL phase.  This mode, usually
denoted as $H$, can be interpreted as the Anderson-Bogoliubov
mode~\cite{AB} as encountered in neutral superfluids such as
superfluid $^3$He and superfluid neutron matter.  In this mode,
fluctuations in the phase of the superfluid order parameter are
coupled with baryon density.  Consequently, this mode behaves as a
longitudinal compression mode; at zero temperature, its velocity is
equal to the velocity of first sound.  In the BCS framework, this mode
can be viewed as a coherent superposition of Cooper pair excitations
with total momentum $\vec{q}$, and hence it does not break up phase
coherence in the superfluid state.  We remark that as far as
excitations associated with baryon number are concerned, symmetric CFL
quark matter behaves like a neutral superfluid even though each quark
pair carries electric and color charge.

     As we have already seen, the $H$ propagator has a massless pole.
Correspondingly, the spectral function naturally has a
$\delta$-function peak.  In the $H$ channel the mixing effect makes
only a little difference in the spectral function.  As we shall see in
the next subsection, however, it plays an important role in the vector
channel.

\begin{figure}
 \includegraphics[width=8cm]{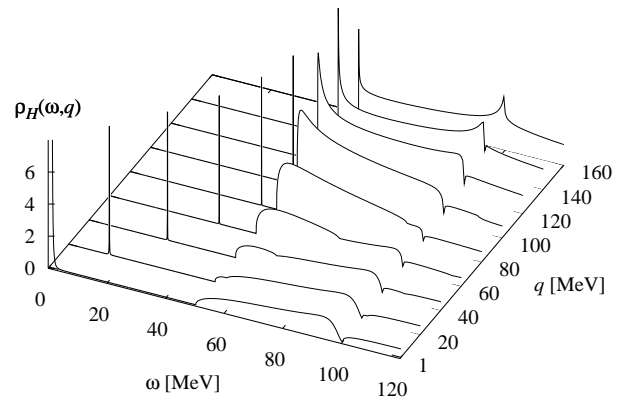}
 \caption{Spectral function in the $H$ channel at zero temperature.}
 \label{fig:chi_spect_zero}
\end{figure}

     From Fig.~\ref{fig:chi_spect_zero}, we can clearly observe that a
massless $H$ mode appears as a $\delta$-function peak (strictly
speaking there is a small width coming from a finite regulator for
numerical computations) and that the continuum starts at the threshold
$\omega=2\Delta$ above which individual pair excitations are
kinematically allowed.  The $H$ mode always lies below the threshold;
if lying above the threshold, it would be superseded by an excited
Cooper pair, which eventually decay into two octet quark
quasiparticles having an energy gap $\Delta$.  We remark that there
appears a marked structure in the continuum around $\omega=4\Delta$.
This energy corresponds to a threshold above which individual pair
excitations of singlet quarks are allowed to occur.

     We can also observe that two spectral peaks near $\omega=2\Delta$
and $\omega=4\Delta$ are appreciable for $q\gtrsim120\MeV$ and that as
the momentum $q$ increases, they become sharper and sharper.  This
spectral enhancement of the continuum around $\omega=2\Delta$ is
analogous to the spectral enhancement as intensely argued in the
chiral system of $\pi$ and $\sigma$ mesons in connection with chiral
restoration~\cite{Hatsuda:1985eb,Yokokawa:2002pw}.  In the present
case no phase transition occurs with increasing $q$, but it seems
reasonable to consider that a superfluid behaves more like a normal
fluid for larger $q$, which will be illustrated in the next
subsection.  In other words, the medium in which $H$ propagates
undergoes continuous changes from superfluid type to normal fluid type
with increasing $q$.

     This continuum structure characterized by the two distinct gaps for 
octet and singlet quarks is a unique feature of the CFL phase, which can be 
seen neither in superfluid $^3$He nor in the 2SC phase, and provides useful
information for identifying the CFL phase.

     We remark that the two peaks near $\omega=2\Delta$ and
$\omega=4\Delta$ do not move substantially until $q$ exceeds $2\mu_r$.
This is because the momenta of two quark quasiparticles, $p$ and
$|\vec{p}+\vec{q}|$, can sit on the Fermi surface simultaneously as
long as $q<2\mu_r$.  We have confirmed that the peaks move upward for
$q>2\mu_r$, although we will not present the results here.

\begin{figure}
 \includegraphics[width=8cm]{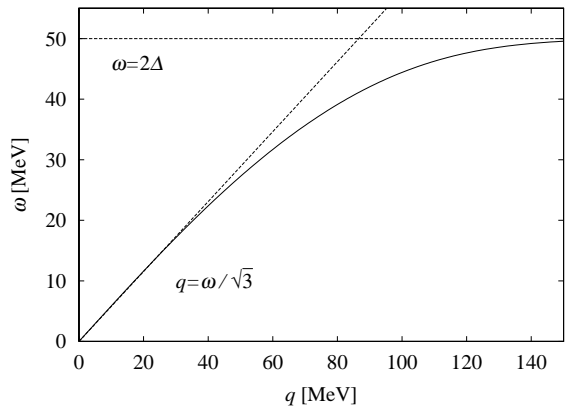}
 \caption{The energy dispersion relation of $H$ as a function of $q$
 at zero temperature.}
 \label{fig:dispersion}
\end{figure}

     Another important property that can be observed from
Fig.~\ref{fig:chi_spect_zero} is the excitation energy of $H$ as a
function of $q$, i.e., the energy dispersion relation of $H$.  The
excitation energy of $H$ is proportional to $q$ at small $q$, and
monotonically increases with $q$ in such a way as to approach the
threshold $\omega=2\Delta$ asymptotically.  This behavior is
consistent with the QCD analysis performed by Zarembo in weak
coupling~\cite{Zarembo:2000pj}.  For further clarity, we plot our
numerical results for the energy dispersion relation in
Fig.~\ref{fig:dispersion}.  In the numerical calculations of the
energy of $H$, we have ignored corrections by the mixing effect, which
are in fact small at low energies and vanish at $\omega=0$.  The
dispersion relation shown in Fig.~\ref{fig:dispersion} thus fulfills
$6/G-\Pi_H(\omega,q)=0$.

     The earlier analyses based on the chiral effective
Lagrangian~\cite{Casalbuoni:1999wu} are valid for $\omega$ and $q$
much lower than $2\Delta$.  The present NJL model calculation is
expected to reproduce the results from the low energy effective
theories.  As can be seen from Fig.~\ref{fig:dispersion}, the speed of
$H$ is $v_H=1/\sqrt{3}$, which is in agreement with the generally
accepted value in the chiral effective Lagrangian approach.  This is
equal to the velocity of first sound, which is consistent with the
fact that the restoring force of this mode is purely kinematic.

     More technically, as shown in Ref.~\cite{Rho:1999xf}, the $H$
velocity of $1/\sqrt{3}$ originates from the dimensionality (three
spatial directions and one temporal direction).  In the rest of this
subsection we shall calculate the $H$ decay constant, $f_H$, by
following the line of the standard NJL calculation of $f_\pi$.  Then,
we shall consider how $v_H=1/\sqrt{3}$ results from the ratio between
the decay constants in the spatial direction, $f^{\rm s}_H$, and in
the temporal direction, $f^{\rm t}_H$.

\begin{figure}
 \includegraphics[width=6cm]{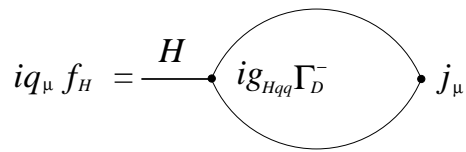}
 \caption{Diagram contributing to $f_H$.}
 \label{fig:fh}
\end{figure}

     We can obtain $f_H$ from the loop diagram shown in
Fig.~\ref{fig:fh} as in the case of $f_\pi$ in chiral dynamics.  The
diagram can be expressed as
\begin{align}
 & iq_\mu f_H \notag \\
 &= -\frac{1}{2}\!\int^T \!\!\frac{d^4 p}{(2\pi)^4}\text{tr}
  \frac{1}{2\sqrt{6}}\Gamma^\mu_V S(p_0\!+\!q_0,\vec{p}\!+\!\vec{q}) 
   ig_{Hqq}\Gamma^-_D S(p_0,\vec{p}),
  \label{eq:fh}
\end{align}
where the factor $1/2$ allows for the Nambu-Gor'kov counting, and the
normalization factor $1/2\sqrt6$ in front of $\Gamma^\mu_V$ is
determined so as to satisfy the proper current algebra.  The $H$-quark
coupling constant, $g_{Hqq}$, can be derived from the residue of the
$H$ propagator; the expression for $g_{Hqq}$ is listed in
Appendix~\ref{app:explicit}.  Then, we find
$f_H =(-1/\sqrt{6})\Delta g_{Hqq}^{-1}$.  [The sign of $f_H$ is
irrelevant to observables because only its square, $f_H^2$, appears in
low energy effective theories~\cite{Casalbuoni:1999wu}.]  After all we
obtain the Goldberger-Treiman relation for quarks in the CFL phase,
\begin{equation}
 (f_H \cdot g_{Hqq})^2 = \frac{1}{6}|\Delta|^2 \,,
\label{eq:goldberger}
\end{equation}
which was also derived in Ref.~\cite{Rho:1999xf} from QCD in weak
coupling.  We note that $g_{Hqq}$ remains finite and thus $f_H$
vanishes in the limit $\Delta\to0$ as it should.

     When $q^\mu$ is infinitesimal and the temperature is much lower
than the quark chemical potential, it is sufficient to limit our
deliberation here to the particle part of the quark propagators in
Eq.~(\ref{eq:fh}).  Then, one can show that the integrand of
Eq.~(\ref{eq:fh}) is proportional to $q_0$ in the limit, $\vec{q}=0$
and $q_0\simeq0$, relevant for $f_H^{\rm t}$, while this $q_0$ is
replaced by $(\widehat{p}\cdot\vec{q})\widehat{p}_i$ in the limit,
$q_0=0$ and $q_i\simeq0$, relevant for $f_H^{\rm s}$.  By noting the
three-dimensional rotational symmetry, one can rewrite
$\int d^3p\,(\widehat{p}\cdot\vec{q})\widehat{p}_i\dots$ as
$(q_i/3)\int d^3p\dots$.  Consequently, the ratio
$f_H^{\rm s}/f_H^{\rm t}$ amounts to
$g_{Hqq}^{\rm s}/3g_{Hqq}^{\rm t}$.  By using this result, the
relation $v_H^2=(f_H^{\rm s}/f_H^{\rm t})^2$ from the chiral effective
Lagrangian, and the Goldberger-Treiman relation~(\ref{eq:goldberger}),
we finally obtain
\begin{equation}
 v_H^2=\frac{1}{3}.
\end{equation}
From the derivation of this result, it is clear that $1/3$ comes from
the spatial dimensionality.

     Let us next derive the expression for $f_H$ in the NJL model.
Using the Goldberger-Treiman relation~(\ref{eq:goldberger}) and the
approximate zero-temperature expression~(\ref{eq:g_t}) for
$g_{Hqq}^{\rm t}$ as derived in Appendix~\ref{app:explicit}, we can
express the temporal decay constant at zero temperature as
\begin{equation}
 (f_H^\text{t})^2 \simeq \frac{3\mu_r^2}{8\pi^2} \,.
\end{equation}
This result is the same as the one derived from QCD in weak
coupling~\cite{Casalbuoni:1999wu}.  Here we emphasize that the
mean-field quark propagator in the NJL model takes the same form as
that in QCD in weak coupling.  As long as diagrams composed of quarks
are concerned, therefore, the NJL model and QCD yield essentially the
same results within the mean-field approximation.  This fact can
explain why the NJL model calculations of $f_\pi$ for CFL pions have
been found to be close to the QCD results in
Refs.~\cite{Buballa:2004sx,Forbes:2004ww}.  We remark that the same
kind of agreement between the mean-field NJL and QCD calculations can
be also found in the parameters characterizing the Ginzburg-Landau
free energy~\cite{Fukushima:2004zq}.

     The results for $f_H$ and $v_H$ obtained in this subsection imply
that the NJL model calculations should encompass the results from
chiral effective Lagrangian approach and even the QCD results in the
mean-field approximation.

%---   Zero sound   ---%

\subsection{Zero sound}
\label{sec:zero_sound}

     We turn to the analysis of the spectral function in the vector
channel at zero temperature.  There are two important features to be
clarified in this subsection: the appearance of zero sound and the
mixing effect between the $H$ and vector channels.  Zero sound is
distinct from the $H$ mode in the sense that it coherently involves a
number of particle-hole excitations rather than pair excitations.  We
first articulate how zero sound emerges in the vector channel in a
normal fluid of quark matter.  We then consider zero sound in the CFL
phase where the mixing effect plays an important role.

% In the normal phase

\subsubsection{In the normal phase}

\begin{figure}
 \includegraphics[width=8cm]{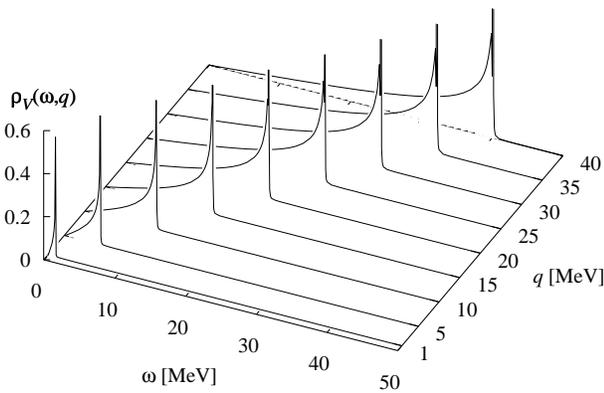}
 \caption{Spectral function in the vector channel, calculated in the
  normal phase at zero temperature.}
 \label{fig:normal}
\end{figure}

     In a normal fluid, where $\Delta=0$ and the baryon number is
conserved, no mixing between the $H$ and vector channels arises, and
we can deal with the vector channel alone.  Our primary results are
illustrated in Fig.~\ref{fig:normal} where the spectral function in
the vector channel is plotted.  In calculating the vector spectral
function in the normal phase, we have fixed all the parameters at the
values adopted in the previous section except $\Delta$, which is set
to be zero.  We can observe from the figure that the continuum ranges
$0 \leq \omega \leq q$ and that an undamped mode (zero sound) stands
in the vicinity of the continuum, although it is rather hard to see in
the plot that the $\delta$-function peak is away from the continuum.

     The continuum in the spectral function stems from individual
excitations of quark particle-hole pairs.  The continuum corresponds
to the space-like region $\omega<q$.  In the present NJL model, the
energy of a quasiparticle of momentum $\vec{k}$ is $k$ as in the case
of a massless free quark gas.  For a real pair of a particle of
momentum $\vec{k}_1$ and a hole of momentum $\vec{k}_2$, therefore, we
can calculate the excitation energy as
$\omega=(k_1-\mu_r)+(\mu_r-k_2)\leq q$, where
$\vec{q}=\vec{k}_2-\vec{k}_1$ is the momentum of the pair.

     If zero sound lies in the continuum, it would suffer Landau
damping.  In this case, zero sound would be absorbed by a
quasiparticle with momentum below the Fermi surface, which
subsequently would be scattered into a state above the Fermi
surface. This process corresponds to the decay of zero sound into a
real particle-hole pair, which is kinematically allowed only when
$\omega<q$.  As we shall show explicitly, however, the calculated
dispersion relation lies above the continuum.

     We now proceed to determine the velocity of zero sound by
following a line of the conventional field-theoretical
argument~\cite{abrikosov}.  Let us focus on the expression for
$\Pi_V$, which is given by Eq.~(\ref{eq:exp_v}) in the limit of
$\Delta\to0$.  At zero temperature the quasiparticle distribution
function vanishes.   The dominant term that remains in $\Pi_V$ is then
\begin{equation}
 \begin{split}
 & \Pi_V(\omega,\vec{q})\simeq 18\int\frac{d^3 p}{(2\pi)^3}
  \Bigl[\Theta(p\!-\!\mu_r)\Theta(\mu_r\!-\!|\vec{p}+\vec{q}|) \\
 & +\Theta(\mu_r\!-\!p)\Theta(|\vec{p}\!+\!\vec{q}|\!-\!\mu_r)\Bigr]
  \frac{\bigl||\vec{p}\!+\!\vec{q}|\!-\!\mu_r\bigr|+|p\!-\!\mu_r|}
 {\Bigl(\bigl||\vec{p}\!+\!\vec{q}|\!-\!\mu_r\bigr|\!+\!|p\!-\!\mu_r|
   \Bigr)^2 \!\!-\omega^2}.
 \end{split}
  \label{eq:pivnf}
\end{equation}
Here $\Theta$ denotes the Heviside's step function.  The first
(second) term in the square brackets corresponds to a virtual
excitation of a particle (hole) with momentum $\vec{p}$ and a hole
(particle) with momentum $\vec{p}+\vec{q}$.  Because of these terms,
the dominant contribution to $\Pi_V$ comes from a regime that
satisfies $p\simeq\mu_r$ and $\omega, q\ll\mu_r$.  We can thus use the
approximation, $|\vec{p}+\vec{q}|\simeq p+\widehat{p}\cdot\vec{q}$.
Eventually, $\Pi_V$ can be estimated as
\begin{equation}
 \Pi_V(\omega,\vec{q})\simeq \frac{18\mu_r^2 q^2}{(2\pi)^3}
  \int d\Omega \frac{\bigl(\vec{n}\cdot\widehat{q}\bigr)^2}
  {q^2\bigl(\vec{n}\cdot\widehat{q}\bigr)^2-\omega^2} \,,
 \label{eq:pivnff}
\end{equation}
where $\int d\Omega=2\pi\int_{-1}^1 d\cos\theta$ and
$\vec{n}\cdot\widehat{q}=\cos\theta$.

     By defining the zero sound velocity as $v_0=\omega/q$ and using
Eq.~(\ref{eq:pivnff}), we can rewrite the zero sound dispersion
relation, $1/G_V+\Pi_V(\omega,q)=0$, as
\begin{equation}
 \frac{v_0}{2}\ln\biggl|\frac{v_0+1}{v_0-1}\biggr| - 1
  = \frac{1}{G_V}\cdot\frac{\pi^2}{9\mu_r^2} \;.
\label{eq:zero_sound}
\end{equation}
This agrees with the conventional form of the nonrelativistic zero
sound equation (see Eq.~(2.26) of Ref.~\cite{abrikosov}).  Since the
right side of Eq.~(\ref{eq:zero_sound}) is $2.304$ for our parameter
choice, we obtain $v_0=1.0028$ as a solution to
Eq.~(\ref{eq:zero_sound}).  Our numerical results for $q=40\MeV$, on
the other hand, has a $\delta$-function peak at $\omega=40.11\MeV$,
from which we can derive the speed as $v_0=40.11/40=1.0028$.  This
value is in excellent agreement with the analytic estimate.

     It is to be noted that the velocity of zero sound is slightly
larger than unity.  This uncausal behavior is due to our negligence of
the longitudinal spatial components of the bubble diagrams in the
vector channel, which contain a term proportional to the cut-off
squared $\Lambda^2$.  We have confirmed that our calculations, if
including the longitudinal spatial components, would give rise to zero
sound of velocity below unity, which is damped in the continuum.
However, it is more important to note that the Fermi velocity is unity
since our model ignores the quark wave-function renormalization.
Undamped zero sound, if present, would thus be automatically
superluminal.  In the high density limit of QCD in which one-gluon
exchange dominates the quark-quark interactions, on the other hand,
the bubble diagram in the vector channel yields a continuum in the
spectral function at $\omega<v_{\rm F}q<q$, where $v_{\rm F}$ is the
Fermi velocity, and an undamped zero sound mode of velocity less than
unity but larger than $v_{\rm F}$~\cite{Baym:1975va}.  Our
calculations, therefore, are successful in reproducing the weak
coupling behavior of zero sound predicted from QCD, except for the
slightly superluminal velocity of sound.  We thus believe that our
framework gives a sufficient basis in analyzing the properties of zero
sound in the CFL phase, as will be discussed below.

% In the CFL phase

\subsubsection{In the CFL phase}

     Let us now examine zero sound and $H$-vector mixing in the CFL
phase.  We first present our numerical results for the spectral
function in the vector channel and then discuss the underlying physics
of the observed spectral suppression at small $q$ and attenuation of
zero sound.

\begin{figure}
\includegraphics[width=8cm]{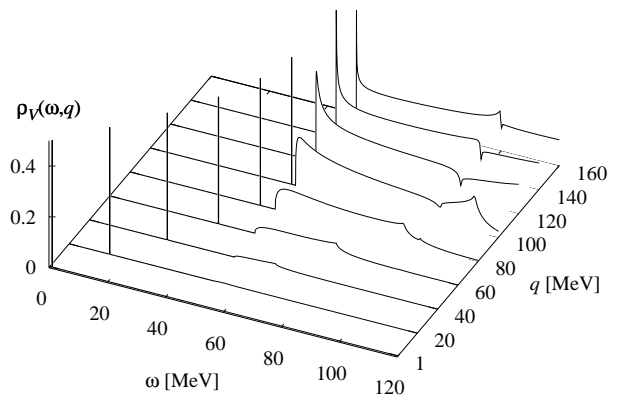}
\caption{Spectral function in the vector channel, calculated in the CFL
 phase at zero temperature.}
\label{fig:v_spect_zero}
\end{figure}

\begin{figure}
 \includegraphics[width=8cm]{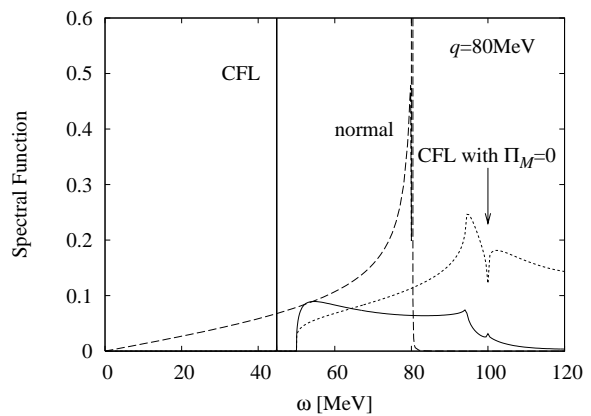}
 \caption{Spectral functions for $q=80\MeV$ in the vector channel,
 calculated in the normal phase (dashed line), in the CFL phase
 without mixing (dotted line), and in the CFL phase with mixing (solid
 line).  The zero sound peak at $\omega\simeq80\MeV$ in the normal
 phase shifts to $\omega\simeq94\MeV$ with an attenuation width in the
 CFL phase.}
 \label{fig:three_spect}
\end{figure}

     We plot the spectral function in the vector channel in
Fig.~\ref{fig:v_spect_zero}, which is to be compared with
Fig.~\ref{fig:normal}.  The spectral functions displayed in
Figs.~\ref{fig:normal} and \ref{fig:v_spect_zero} are markedly
different in two respects: In the CFL phase a $\delta$-function peak
corresponding to an $H$ phonon in the space-like ($\omega<q$) region
appears, and a peak  corresponding to zero sound, located at
$\omega\simeq78\MeV$ for $q=60\MeV$, at $\omega\simeq94\MeV$ for
$q=80\MeV$, and at $\omega\simeq111\MeV$ for $q=100\MeV$, has a width
in the continuum that extends above $\omega=2\Delta$.  As a result of
the $H$-vector mixing, the $H$ phonon peak and continuum similar to
the ones in the $H$ channel as shown in Fig.~\ref{fig:chi_spect_zero}
do appear in the vector channel, but have a vanishingly small spectral
weight at $q\simeq0$.  On the other hand, the observed width of the
zero sound peak indicates the attenuation of zero sound.  In order to
see how the attenuation and the mixing effect affect the spectral
function, we plot in Fig.~\ref{fig:three_spect} three spectral
functions calculated in the normal phase, in the CFL phase by ignoring
the mixing effect, and in the CFL phase by including the mixing
effect.

     First, we discuss the strong spectral suppression in the vector
channel at $q\simeq0$ and the underlying mixing effect between the $H$
and vector channels.  In the long wavelength limit $q=0$, a clear
distinction between particle and hole excitations is lost in the
superfluid medium, and only the $H$ (particle-particle) channel
contributes to the spectral weight.  For $q$ far larger than $\Delta$,
on the other hand, the medium behaves more like a normal fluid rather
than a superfluid, and the vector (particle-hole) channel exhibits a
similar spectral function to that in the normal phase.

     We can explicitly show the spectral suppression by noting that
the bubble diagrams in the limit $q\to0$ satisfy the relations,
\begin{align}
 \Pi_H(\omega,0) & = \Pi_H(0,0)
   + \frac{\omega^2}{4\Delta^2}\Pi_V(\omega,0) \notag\\
  & = \frac{6}{G} + \frac{\omega^2}{4\Delta^2}\Pi_V(\omega,0), \\
 \Pi_M(\omega,0) & = -\frac{i\omega}{2\Delta}\Pi_V(\omega,0) \,.
\end{align}
As a result of these relations, the vector component of the propagator
matrix (\ref{eq:ret_prop}) reduces to $-G_V$, which is exactly the
tree-level value (the cross diagram in Fig.~\ref{fig:diagram}).  This
indicates that the vector channel no longer contains the quark loops
responsible for collective and individual excitations, leading to a
vanishing spectral function.  In other words, any excitations with
small momenta in the vector channel tend to be absorbed into the $H$
channel in the superfluid medium.  As $q$ becomes larger, the vector
spectral function grows in the medium in which particle and hole
excitations are no longer indistinguishable, leading to a zero sound
peak.

     This difference in the vector channel at small $q$ between the
normal and CFL phases could serve as a possible diagnosis of
superfluidity of quark matter if the spectral function in the vector
channel could be measured in the lattice QCD simulation.  Quark matter
could be considered to be in the normal phase when zero sound is found
at small $q$, while in the superfluid phase when strong spectral
suppression or weak mixture with an $H$ phonon is observed.  One could
distinguish between zero sound and an $H$ phonon from the fact that
the $H$ phonon velocity is smaller than the zero sound velocity by a
factor of $\sim1/\sqrt{3}$ except for Fermi-liquid corrections.  In
fact, zero sound has been found in the lattice simulation based on a
QCD-like model~\cite{Hands:2003dh}.  If the simulation involves a
superfluid phase, the vector channel measurement would provide a clear
signature of superfluidity.

\begin{figure}
 \includegraphics[width=8cm]{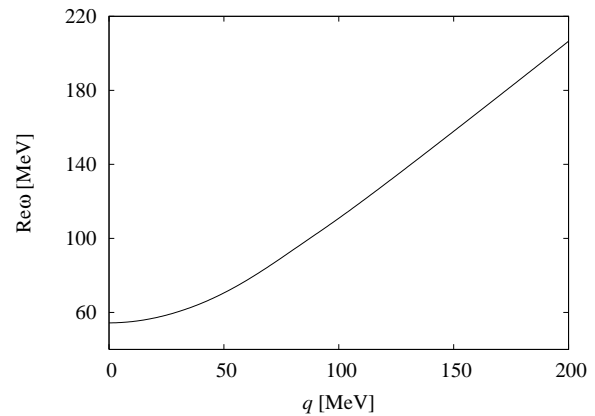}
 \caption{The real part of the zero sound pole in the CFL phase at
  zero temperature.}
 \label{fig:v_disp}
\end{figure}

\begin{figure}
 \includegraphics[width=8cm]{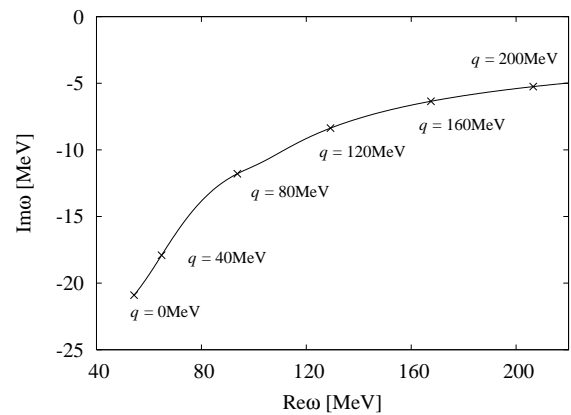}
 \caption{The pole position of zero sound in the CFL phase at zero
  temperature.  The curve is slightly distorted when it passes
  through the threshold energy $\mathrm{Re} \omega=100\MeV$ for
  individual pair excitations of singlet quarks.}
 \label{fig:v_pole}
\end{figure}

     We next discuss the attenuation of zero sound in the continuum.
In order to extract detailed information on zero sound, we search for
a pole of the vector propagator in the complex energy plane, which is
responsible for a weak peak in the spectral function in the vector
channel.  For simplicity we ignore the $H$-vector mixing and the
antiparticle contribution in locating the complex pole position.  This
kind of pole search has been performed in the context of chiral
restoration in a hot and dense medium~\cite{Yokokawa:2002pw} and of a
phase transition from the normal to the 2SC
phase~\cite{Kitazawa:2001ft}.  Appendix~\ref{app:analytic} is devoted
to exhibiting how to perform the analytic continuation into the second
Riemann sheet where the attenuated mode would be found.

     In Figs.~\ref{fig:v_disp} and \ref{fig:v_pole} we plot the pole
position, $\omega$, as a function of $q$.  We can observe from
Fig.~\ref{fig:v_disp} that the real part of $\omega$ almost linearly
increases with $q$ once $q$ becomes larger than $\sim2\Delta$.  Since
there is no quark excitation at low $\omega$ due to a nonzero pairing
gap, zero sound, a coherent superposition of particle-hole pairs, can
only emerge in the continuum where quasiparticle modes are allowed to
excite.  This explains why the energy of zero sound has a gap and
stays slightly larger than $2\Delta$ for $q\to0$.

     Figure~\ref{fig:v_pole} illustrates how the pole moves in the
complex energy plane with increasing $q$.  The width of zero sound
attenuation decreases with $q$.  This attenuation occurs in the
superfluid medium in which zero sound is absorbed by a Cooper pair,
which subsequently breaks up into two quasiparticles.  Consequently,
the decrease in the width with increasing $q$ is consistent with the
tendency that the medium behaves more like a normal fluid for
elementary excitations with larger $q$.  Recall that zero sound in the
normal phase is undamped in the present analysis.

     We conclude this section by summarizing the zero-temperature
results for an $H$ phonon and zero sound in the CFL phase.  $H$
phonons are only low-lying longitudinal excitations and survive as
long as $\omega<2\Delta$, while zero sound manifests itself as an
attenuated mode for $\omega>2\Delta$.  The former is consistent with
the earlier results from low energy effective theories.  The latter is
the point that we first clarified in the context of superfluid quark
matter.  The presence of the phase and zero sound modes in different
regimes of $\omega$ can also be seen in superfluid
$^3$He~\cite{Wolfle}.  We also discovered a strong spectral
suppression in the vector channel at small $q$.  We remark in passing
that all these properties of the CFL state are robust even if the
longitudinal spatial components of the bubble diagrams in the vector
channel are taken into account.

%%%%%%%%%%   Finite Temperature Results   %%%%%%%%%%

\section{finite temperature results}
\label{sec:finite}

     In this section we examine thermal effects on collective
excitations in the collisionless limit.  The collisions between
quasiparticles, which are expected to play a role in modifying the
properties of the collective modes at finite temperatures, will be
addressed in our future study.  At finite temperatures, as we shall
see, the $H$ modes are Landau damped by thermally excited quark
quasiparticles, while the behavior of zero sound is essentially the
same as that obtained at zero temperature.  Throughout this section we
limit ourselves to the case at $T=0.8T_{\rm c}=14.52\MeV$, which is
high enough for us to perceive the temperature effect in the $H$
channel, while not being in the immediate vicinity of the critical
point.  At this temperature, the pairing gap and the effective
chemical potential can be calculated as $\Delta=16.4\MeV$ and
$\mu_r=435.3\MeV$.  The temperature dependence of $\mu_r$ turns out to
be tiny; even just above $T_{\rm c}$, $\mu_r$ is greater than the zero
temperature value only by 2\%.

     We first consider zero sound, which, at zero temperature, appears
as a low-lying mode in the normal phase and as an attenuated mode
above $2\Delta$ in the CFL phase, as we have seen in
Sec.~\ref{sec:zero_sound}.  The behavior of zero sound at
$T=0.8T_{\rm c}$ and at $T=0$ is almost indistinguishable.  This is
because $T=0.8T_{\rm c}=14.52\MeV$ is \textit{low} compared with the
Fermi temperature corresponding to the quark chemical potential
$\sim500\MeV$.  In the normal phase, unless the temperature is
considerably higher than $T_{\rm c}$, no drastic change occurs in the
behavior of zero sound from that shown in Fig.~\ref{fig:normal}.

\begin{figure}
 \includegraphics[width=8cm]{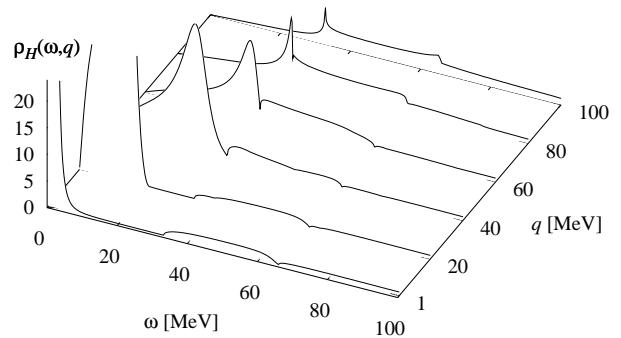}
 \caption{Spectral function in the $H$ channel, calculated in the CFL
  phase at $T=0.8T_{\rm c}$.}
 \label{fig:chi_spect_finite}
\end{figure}

     In contrast to the case of zero sound, $H$ modes in the CFL phase
are damped only at nonzero temperatures.  In the case in which
thermally excited quark quasiparticles are present, the $H$ modes
suffer Landau damping.  This process can be viewed as the upscattering
of a quasiparticle by absorbing a $H$ phonon, which is characterized
by the nonvanishing factor $[n_{\rm F}^{-} - n_{\rm F}^{-\prime}]$ in
Eq.~(\ref{eq:exp_h}).  The similar damping process of the phase modes
has been considered in the context of neutral BCS
superfluids~\cite{TDGL}.  The Landau damping of the $H$ modes can be
observed from Fig.~\ref{fig:chi_spect_finite} in which the spectral
function in the $H$ channel is plotted at $T=0.8T_{\rm c}$.  We find
that the $H$ modes have a finite decay width in contrast to the zero
temperature case as shown in Fig.~\ref{fig:chi_spect_zero}.  The
threshold for pair excitations is located at
$\omega=2\Delta=32.8\MeV$.  We also note that as a result of the
Landau damping, the spectral function has a continuum below the $H$
peak.

     Detailed analysis shows that the velocity of the $H$ mode at
$T=0.8T_{\rm c}$ is $v_H=0.554$ which is slightly smaller than
$1/\sqrt3$.  This is because the superfluid part rather than the whole
system participates in the phase fluctuations.

     Now, we briefly discuss the possible relevance of our results to
the transport properties in the CFL phase.  In the $H$ channel, by
setting $\Pi_M=0$ for simplicity, we can find a pole on the imaginary
axis in the complex energy plane.  This pure imaginary pole
corresponds to a diffusive mode that does not propagate at all.  We
identify this imaginary pole as responsible for the low-lying
continuum due to the Landau damping.

\begin{figure}
 \includegraphics[width=8cm]{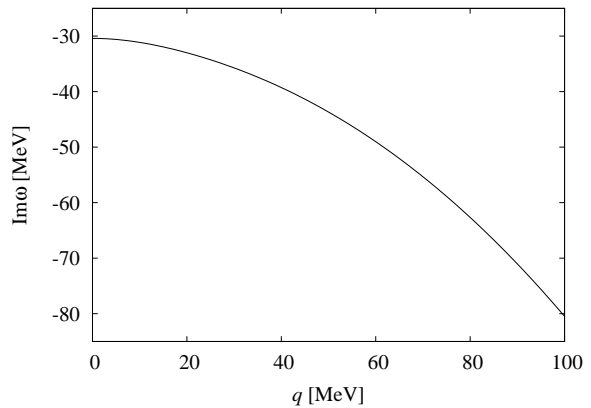}
 \caption{The pole position of the diffusive mode responsible for the
  Landau damping in the CFL phase at $T=0.8T_{\rm c}$.}
 \label{fig:landau}
\end{figure}

     In Fig.~\ref{fig:landau} we plot the pole position as a function
of the momentum $q$.  The result can be parametrized as
$\mathrm{Im}\,\omega=-30.9-5.11\!\times\! 10^{-3}q^2\:\text{[MeV]}$.
This result suggests that $H$ modes in CFL quark matter suffer
diffusion with the diffusion constant
$5.11\!\times\!10^{-3}\,\text{MeV}^{-1}$ in addition to absorption.
This diffusion coefficient arises even in the collisionless limit, but
has to be modified by collisions between quasiparticles, which are
more relevant at higher temperatures.

     Although description of collective modes in the hydrodynamic
regime is beyond the scope of the present paper, we shall comment on
what is expected in the hydrodynamic regime.  In the normal phase, as
in a normal liquid $^3$He, zero sound in the collisionless regime
should continuously transform to first sound in the hydrodynamic
regime.  In the superfluid medium, as discussed in Ref.~\cite{lee59}
and also pointed out in Ref.~\cite{Fukushima:2004bj}, the mode
corresponding to phase oscillations in the collisionless regime should
change into second sound in the hydrodynamic regime.  It would be an
interesting future problem to describe such continuous transitions in
the nature of sound modes within the framework of kinetic theory.

%%%%%%%%%%   Concluding Remarks   %%%%%%%%%%

\section{concluding remarks}
\label{sec:conclusion}

     In this paper we have studied the longitudinal spectral functions
in the $H$ and vector channels in a superfluid of CFL quark matter.
We have elucidated the zero-temperature properties of the vector
channel: the strong spectral suppression at small momenta due to the
$H$-vector mixing and the zero sound attenuation above the energy
threshold for pair excitations.  The important implication of these
results is that the spectral function in the vector channel is
sensitive to the state of quark matter.  If lattice QCD simulation at
finite density is feasible in the future, the vector channel
measurements could be of great use for identifying superfluidity of
quark matter.

     Extension of the present analysis to the case of nonzero quark
masses is desirable for the description of elementary excitations in
realistic situations.  In this case, as discussed in
Sec.~\ref{sec:cooper}, the possible coupling of phase and baryon
density fluctuations with electric and/or color charge density
fluctuations could change the properties of the longitudinal
collective excitations drastically.  In the presence of gapless quark
modes as in the gapless CFL phase~\cite{Alford:2003fq}, the continuum
of the quark excitations and hence the attenuation of the collective
modes would be enhanced.

     In the present analysis we have utilized the NJL model for quark
matter.  The NJL model description turns out to reproduce the results
for $H$ phonons known from low-energy effective theories and even from
QCD in the mean-field approximation.  In the collisionless regime,
however, the longitudinal spectral functions remain to be clarified
from QCD even in weak coupling.  Furthermore, collective modes in the
CFL phase have yet to be examined in the hydrodynamic regime.  It
would be instructive to describe low-lying excitations of CFL quark
matter in both regimes by using a relativistic version of the Landau
theory of Fermi liquids~\cite{Baym:1975va}, which effectively
encompasses the NJL model.

\acknowledgments
     We are grateful to G.~Baym and T.~Hatsuda for valuable comments
on these and related subjects.  K.~F.\ thanks M.~Forbes, C.~Kouvaris,
K.~Rajagopal, and D.~T.~Son for discussions.  This work was partially
supported by the Japan Society for the Promotion of Science for Young
Scientists and by the U.S. Department of Energy (D.O.E.) under
cooperative research agreement \#DF-FC02-94ER40818.

%%%%%%%%%%   Appendix   %%%%%%%%%%

\appendix

%%%%%%%%%%   Evaluation of the Bubble Diagrams   %%%%%%%%%%

\section{evaluation of the bubble diagrams}
\label{app:explicit}

     In this appendix, we present the explicit expressions for the
polarization functions in the $H$, vector, and mixed channels.  We
also provide the calculation of $g_{Hqq}$ which is a derivative of the
polarization with respect to either energy or momentum.  Since the
quark propagator is divided into the particle part and the
antiparticle part as displayed in Eq.~(\ref{eq:propagator}), the
bubble diagram in the $l$ channel ($l=H,V,M$) can be divided into four
parts as
\begin{equation}
 \Pi_l = \Pi_l^{\text{pp}} + \Pi_l^{\text{aa}}
  + \Pi_l^{\text{pa}} + \Pi_l^{\text{ap}},
\end{equation}
where the superscript ``p'' and ``a'' denote the particle and
antiparticle parts of the quark propagator, respectively.

     We begin with the expression for the polarization $\Pi_H$ in the
$H$ channel.  The particle-particle part can be written as
\begin{widetext}
\begin{equation}
 \begin{split}
 & \Pi^{\text{pp}}_H (\omega,\vec{q}) = 2\int^\Lambda\frac{d^3 p}
  {(2\pi)^3}\bigl[1+\widehat{p}\cdot(\widehat{p+q})\bigr]
  \frac{1}{\epsilon^-_\Delta \epsilon^{-\prime}_\Delta} \\
 &\qquad
  \times \Biggl\{\Bigl[\epsilon^-_\Delta \epsilon^{-\prime}_\Delta
  +(p-\mu_r)(|\vec{p}+\vec{q}|-\mu_r)+\Delta^2\Bigr]
  \biggl(\frac{1}{\omega+\epsilon^-_\Delta+\epsilon^{-\prime}_\Delta}
   -\frac{1}{\omega-\epsilon^-_\Delta-\epsilon^{-\prime}_\Delta}
  \biggr) \bigl[1-n^-_{\text{F}}-n^{-\prime}_{\text{F}}\bigr] \\
 &\qquad\qquad
  +\Bigl[\epsilon^-_\Delta \epsilon^{-\prime}_\Delta
  -(p-\mu_r)(|\vec{p}+\vec{q}|-\mu_r) -\Delta^2\Bigr]
  \biggl(\frac{1}{\omega-\epsilon^-_\Delta+\epsilon^{-\prime}_\Delta}
   -\frac{1}{\omega+\epsilon^-_\Delta-\epsilon^{-\prime}_\Delta}
  \biggr) \bigl[n^-_{\text{F}}-n^{-\prime}_{\text{F}}\bigr]\Biggr\} \\
 &\qquad\qquad\qquad
  +\frac{1}{2}(\text{same with }\Delta\to 2\Delta) \,,
 \end{split}
\label{eq:exp_h}
\end{equation}
where we have used the notations for the quark excitation energies,
$\epsilon^-_\Delta=\epsilon^-_\Delta(\vec{p})$ and
$\epsilon^{-\prime}_\Delta=\epsilon^-_\Delta(\vec{p}+\vec{q})$,
and for the Dirac-Fermi distribution functions,
$n^-_{\text{F}}=[e^{\epsilon^-_\Delta/T}+1]^{-1}$ and
$n^{-\prime}_{\text{F}}=[e^{\epsilon^{-\prime}_\Delta/T}+1]^{-1}$.
The antiparticle-antiparticle part, $\Pi^{\text{aa}}_H$, can be
readily obtained from expression~(\ref{eq:exp_h}) by replacing $\mu_r$
by $-\mu_r$.  [Note that by this replacement $\epsilon^-$ is replaced
by $\epsilon^+$.]  One can obtain the expression for the
particle-antiparticle part $\Pi^{\text{pa}}_H$ by replacing, in
Eq.~(\ref{eq:exp_h}), $(p-\mu_r)$ by $-(p+\mu_r)$, $\epsilon^-_\Delta$
by $\epsilon^+_\Delta$, $n^-_{\text{F}}$ by $n^+_{\text{F}}$, and
$1+\widehat{p}\cdot(\widehat{p+q})$ by
$1-\widehat{p}\cdot(\widehat{p+q})$.  The antiparticle-particle part,
$\Pi^{\text{ap}}_H$, likewise obtains.

     The particle-particle part of the bubble diagram in the vector
channel reads
\begin{equation}
 \begin{split}
 & \Pi^{\text{pp}}_V (\omega,\vec{q}) = 2\int^\Lambda\frac{d^3 p}
  {(2\pi)^3}\bigl[1+\widehat{p}\cdot(\widehat{p+q})\bigr]
  \frac{1}{\epsilon^-_\Delta \epsilon^{-\prime}_\Delta} \\
 &\qquad
  \times \Biggl\{\Bigl[\epsilon^-_\Delta \epsilon^{-\prime}_\Delta
  -(p-\mu_r)(|\vec{p}+\vec{q}|-\mu_r)+\Delta^2\Bigr]
  \biggl(\frac{1}{\omega+\epsilon^-_\Delta+\epsilon^{-\prime}_\Delta}
   -\frac{1}{\omega-\epsilon^-_\Delta-\epsilon^{-\prime}_\Delta}
  \biggr) \bigl[1-n^-_{\text{F}}-n^{-\prime}_{\text{F}}\bigr] \\
 &\qquad\qquad
  +\Bigl[\epsilon^-_\Delta \epsilon^{-\prime}_\Delta
  +(p-\mu_r)(|\vec{p}+\vec{q}|-\mu_r) -\Delta^2\Bigr]
  \biggl(\frac{1}{\omega-\epsilon^-_\Delta+\epsilon^{-\prime}_\Delta}
   -\frac{1}{\omega+\epsilon^-_\Delta-\epsilon^{-\prime}_\Delta}
  \biggr) \bigl[n^-_{\text{F}}-n^{-\prime}_{\text{F}}\bigr]\Biggr\} \\
 &\qquad\qquad\qquad
  +\frac{1}{8}(\text{same with }\Delta\to 2\Delta).
 \end{split}
\label{eq:exp_v}
\end{equation}
\end{widetext}
This is the same as $\Pi_H^{\text{pp}}$ except for the sign in front
of the factor $(p-\mu_r)(|\vec{p}+\vec{q}|-\mu_r)$ and the coefficient
affixed to the singlet contribution.  This change in the sign can be
intuitively understood once the limit of $\Delta\to0$ is taken.  In
this limit, as discussed in Sec.~\ref{sec:zero_sound}, the integrand
is nonzero only when one of the factors $(p-\mu_r)$ and
$(|\vec{p}+\vec{q}|-\mu_r)$ is positive and the other is negative.
This situation corresponds to a simultaneous excitation of a quark
particle above the Fermi surface and a quark hole below the Fermi
surface, rather than pair excitations which play a role in
$\Pi_H^{\text{pp}}$.  In the presence of the energy gap $\Delta$, the
clear distinction between particles and holes is lost since the
quasiparticle state appears as a linear superposition of a particle
state and a hole state.  The other parts, $\Pi^{\text{aa}}_V$,
$\Pi^{\text{pa}}_V$, and $\Pi^{\text{ap}}_V$, obtain in the same way
as in the $H$ channel.

     We remark that $\Pi_V^{\text{pp}}$ is almost independent of the
cut-off $\Lambda$ because the momentum integration around the Fermi
surface contributes dominantly to $\Pi_V^{\text{pp}}$.  For
sufficiently large $p$ to satisfy $|\vec{p}+\vec{q}|>\mu_r$, a hole is
not allowed to excite at zero temperature, so that the momentum
integration has a natural cut-off.  In contrast, $\Pi_V^{\text{pa}}$
has an appreciable dependence on $\Lambda$ since an antiparticle can
arise at arbitrary $\vec{p}$.  This cut-off dependent contribution is,
however, suppressed by a factor of $q^2/p^2$ because of the
orthogonality between the positive and negative energy states, i.e.,
$\text{tr}\Lambda_{p+q}^\pm \Lambda_p^\mp
 =1-\widehat{p}\cdot(\widehat{p+q})$.  If we take account of the
spatial component of the polarization, $\Pi_V^{33}$, for instance, a
particular choice of the polarizing direction leads to breaking of the
orthogonality.  As a result the particle-antiparticle contribution of
$\Pi_V^{33}$ has a term proportional to the cut-off squared
$\Lambda^2$, and hence the cut-off dependence of $\Pi_V^{33}$ is not
well controllable in contrast to the case of $\Pi_V^{00}$ considered
here.

     We turn to the expression for the polarization $\Pi_M$ in the
mixed channel.  The particle-particle part reads
\begin{widetext}
\begin{equation}
 \begin{split}
 & \Pi^{\text{pp}}_M (\omega,\vec{q}) = -2i\Delta\omega
  \int^\Lambda\frac{d^3 p}{(2\pi)^3}\bigl[1+\widehat{p}\cdot
  (\widehat{p+q})\bigr]
  \frac{1}{\epsilon^-_\Delta \epsilon^{-\prime}_\Delta}
  \Biggl\{\biggl(\frac{1}{\omega+\epsilon^-_\Delta
   +\epsilon^{-\prime}_\Delta} -\frac{1}{\omega-\epsilon^-_\Delta
   -\epsilon^{-\prime}_\Delta}\biggr) \bigl[1-n^-_{\text{F}}
   -n^{-\prime}_{\text{F}}\bigr] \\
 &\qquad
  -\biggl(\frac{1}{\omega-\epsilon^-_\Delta
   +\epsilon^{-\prime}_\Delta}-\frac{1}{\omega+\epsilon^-_\Delta
   -\epsilon^{-\prime}_\Delta}\biggr)
   \bigl[n^-_{\text{F}}-n^{-\prime}_{\text{F}}\bigr]\Biggr\}
  +\frac{1}{4}(\text{same with }\Delta\to 2\Delta),
 \end{split}
\end{equation}
\end{widetext}
and $\Pi_M^{\text{aa}}$, $\Pi_M^{\text{pa}}$, and $\Pi_M^{\text{ap}}$
obtain in the same way as in the $H$ channel.

     From the expression for $\Pi_H$ obtained above, we can calculate
the coupling constant between $H$ and quarks, which is determined by
the residue of the $H$ propagator as
\begin{equation}
 g_{Hqq}^2 = \biggl[\frac{\partial\Pi_H}{\partial q^\mu q_\mu}
  \biggr]^{-1}_{q^\mu q_\mu=m_H^2=0},
\label{eq:gHqq}
\end{equation}
where $m_H$ is the $H$ mass.  We note that since $m_H=0$, the mixed
part $\Pi_M$ does not contribute to $g_{Hqq}$.  Since Lorentz symmetry
is broken in a many particle system, the derivatives in the temporal
and spatial directions lead  to different results, which in turn yield
the velocity of the $H$ mode smaller than unity.

     We can write down the derivative in the temporal direction as
\begin{equation}
 \begin{split}
 & (g^{\text{t}}_{Hqq})^{-2}=
  \frac{\partial\Pi^{\text{pp}}_H(\omega,0)}
  {\partial\omega^2}\Biggr|_{\omega=0} \\
 = &2\int^\Lambda \!\frac{d^3 p}{(2\pi)^3}
  \frac{1-2n_{\text{F}}}{(\epsilon^-_\Delta)^3}
  +\frac{1}{2}\text{(same with $\Delta\to2\Delta$)}
 \end{split}
 \label{eq:gthqq}
\end{equation}
for the particle-particle part.  Here we note that
$\Pi_H^{\text{ap}}(\omega,0)=0$ and that $\Pi_H^{\text{aa}}$ is
suppressed compared with $\Pi_H^{\text{pp}}$ since $\epsilon^-_\Delta$
in $\Pi_H^{\text{pp}}$ is replaced by $\epsilon^+_\Delta$ in
$\Pi_H^{\text{aa}}$.  At zero temperature we can estimate the integral
in Eq.~(\ref{eq:gthqq}), by assuming that it is dominated by momenta
close to the Fermi surface, as
\begin{equation}
 \int\frac{d^3 p}{(2\pi)^3}\frac{1}{(\epsilon^-_\Delta)^3}
  \simeq \frac{\mu_r^2}{2\pi^2}\int_{-\infty}^\infty
  \frac{dp}{(p^2+\Delta^2)^{3/2}} = \frac{\mu_r^2}{\pi^2\Delta^2}.
\end{equation}
We thus obtain the temporal $H$-quark coupling constant at zero
temperature as
\begin{equation}
 (g^{\text{t}}_{Hqq})^{-2} = \frac{9\mu_r^2}{4\pi^2\Delta^2}.
\label{eq:g_t}
\end{equation}

     We next estimate from Eq.\ (\ref{eq:gHqq}) the spatial $H$-quark
coupling constant at zero temperature.  The differentiation in the
spatial direction involved in this estimate is more complicated than
that in the temporal direction.  By expanding the integrand of
$\Pi_H^{\text{pp}}(0,\vec{q})$ with respect to $q$, we find that the
term proportional to $q^2$ is smaller than the term proportional to
$(\widehat{p}\cdot\widehat{q})^2$ by a factor of $\Delta^2/\mu_r^2$.
We thus neglect the former term.  Then, the integrand of $\partial
\Pi_H^{\text{pp}}(0,\vec{q})/\partial q^2$ reduces to
\begin{equation}
 \frac{\Delta^4+(\epsilon^-_\Delta)^2 (\mu_r^2-3\mu_r p+3p^2)
  +\Delta^2 (\mu_r^2-2\mu_r p-2p^2)}{(\epsilon^-_\Delta)^5}.
\end{equation}
Note that the integral with respect to $p$ is dominated by momenta
close to the Fermi surface.  We can thus simplify the integrand by
dropping the $\Delta^4$ term and by setting $p=\mu_r$ in the numerator
as
\begin{equation}
 \simeq \:\frac{\mu_r^2 [(\epsilon^-_\Delta)^2 -3\Delta^2]}
  {(\epsilon^-_\Delta)^5} \,.
\end{equation}
We finally obtain
\begin{equation}
\begin{split}
 & (g^{\text{s}}_{Hqq})^{-2}= -\frac{\partial\Pi^{\text{pp}}_H(0,q)}
  {\partial q^2}\biggl|_{q=0} \\
 =& 2\int^\Lambda \!\frac{d^3 p}{(2\pi)^3}
  \frac{(\widehat{p}\cdot\widehat{q})^2}{(\epsilon^-_\Delta)^3}
 =\frac{1}{3}(g^{\text{t}}_{Hqq})^{-2} \,,
\end{split}
\label{eq:g_s}
\end{equation}
where we have used the relation 
$\int dp\,(\epsilon^-_\Delta)^{-5}
=(2/3\Delta^2)\int dp\,(\epsilon^-_\Delta)^{-3}$, and the factor $1/3$
originates from the average over three spatial directions.

%%%%%%%%%%   Analytic Continuation to the Second Riemann Sheet   %%%%%%%%%%

\section{analytic continuation to the second Riemann sheet}
\label{app:analytic}

\begin{figure}
 \includegraphics[width=8cm]{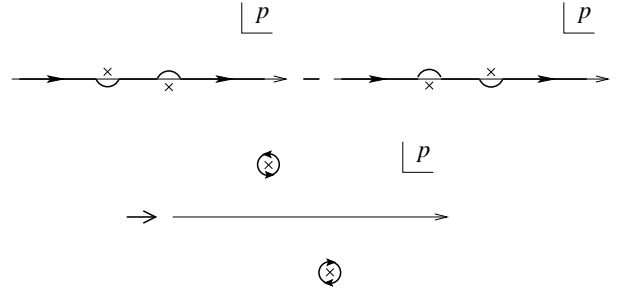}
 \caption{Poles off the real axis in the complex $p$ plane.  We can
  perform the analytic continuation by picking up the positive residue
  of poles in the upper-half plane and the negative residue of poles
  in the lower-half plane.}
 \label{fig:analytic}
\end{figure}

     Here we explain how to perform the analytic continuation to the
second Riemann sheet on which possible unstable states would have a
pole.  In the imaginary-time formalism of finite temperature field
theory, the real-time retarded propagator,
$\boldsymbol{D}^R(\omega,\vec{q})$, can be obtained by the analytic
continuation, $\boldsymbol{D}^R(\omega,\vec{q})
=\boldsymbol{D}(q_0=\omega+i0^+,\vec{q})$.  When $\omega$ is larger
than the threshold for single particle excitations, the propagator has
a finite imaginary part,  which leads to the spectral function having
a continuum.  Then, the propagator and hence the polarization
$\boldsymbol{\Pi}(\omega,\vec{q})$ has a branch cut in the complex
energy plane.

     If we take $\boldsymbol{\Pi}(z=\zeta-i\eta,\vec{q})$ as an
analytically continued function, this polarization in the limit of
$\eta\to0$ is not continuously connected to
$\boldsymbol{\Pi}(\zeta+i0^+,\vec{q})$ in the physical sheet due to a
branch cut in the continuum region.  In order to remedy this
situation, it is convenient to add the discontinuity,
\begin{equation}
 \text{Disc}\boldsymbol{\Pi}(\omega,\vec{q}) 
  = \boldsymbol{\Pi}(\omega+i0^+,\vec{q})
  -\boldsymbol{\Pi}(\omega+i0^-,\vec{q}) \,.
\end{equation}
Once we know the analytic continuation of
$\text{Disc}\boldsymbol{\Pi}(\omega,\vec{q})$, we can locate the pole
in the second Riemann sheet by using
$\boldsymbol{\Pi}(z,\vec{q})+\text{Disc}\boldsymbol{\Pi}(z,\vec{q})$
as a properly analytically continued function.

     In our calculation $\boldsymbol{\Pi}$ is not written as a closed
expression, but, as listed in Appendix~\ref{app:explicit}, an
expression involving the momentum integration with respect to $p$.
For the energy $\omega+i0^+$, the integrand has in general several
poles stemming from
$1/(\omega-\epsilon^-_\Delta-\epsilon^{-\prime}_\Delta)$ and other
similar terms in the complex $p$ plane.  These poles lie either above
or below the real axis as shown in Fig.~\ref{fig:analytic}.  Then, as
is obvious from the figure, $\text{Disc}\boldsymbol{\Pi}$ is nothing
but the contour integrals around the poles.  If a pole is above
(below) the real axis, the contour goes in the positive (negative)
orientation.  Straightforward generalization of this computational
rule to the case of complex $\omega$ yields the analytic
continuation.  For complex $\omega=z$, poles are not necessarily
distributed in the vicinity of the real axis.  Then, we can calculate
$\text{Disc}\boldsymbol{\Pi}(z,\vec{q})$ from the contour integrals
around the poles as shown in the lower part of
Fig.~\ref{fig:analytic}.

%%%%%%%%%%   REFERENCES   %%%%%%%%%%

\end{document}